\documentclass[aps,twocolumn,groupedaddress,showpacs]{revtex4}
\usepackage{ifpdf}
\usepackage{epsfig,amsbsy,bm,amssymb}
\usepackage{hyperref}
\usepackage{graphicx}

\usepackage[english]{babel}
 
\usepackage{natbib}
\newcommand{\isum}%
{\mathop{\hbox{$\displaystyle\sum\kern-13.2pt\int\kern1.5pt$}}}
\renewcommand{\r}{{\bm r}}

\newcommand{\p}{{\bm p}}
  \newcommand{\A}{{\bm A}}

  \newcommand{\ve}{{\bm v}}

\newcommand{\bt}{\begin{tabular}}
\newcommand{\et}{\end{tabular}}

\newcommand{\eref}[1] {(\ref{#1})}
\newcommand{\Eref}[1] {Eq.~(\ref{#1})}

\newcommand{\Fref}[1] {Fig. \ref{#1}}

\newcommand{\br}{\begin{eqnarray*}}
\newcommand{\er}{\end{eqnarray*}}
\newcommand{\ba}{\begin{eqnarray}}
\newcommand{\ea}{\end{eqnarray}}
\newcommand{\be}{\begin{equation}}
\newcommand{\ee}{\end{equation}}

\newcommand{\bp}{\begin{minipage}}
\newcommand{\ep}{\end{minipage}}

\begin{document}

\title {Simple man model in the Heisenberg picture.}

\author{I. A. Ivanov$^{1}$ }
\email{igorivanov@ibs.re.kr}
\author{Kyung Taec Kim$^{1,2}$}
\email{kyungtaec@gist.ac.kr}

\affiliation{$^{1}$Center for Relativistic Laser Science, Institute for
Basic Science, Gwangju, 61005, Korea}

\affiliation{$^{2}$Department of Physics and Photon Science, GIST, Gwangju, 61005, Korea}

\date{\today}

\begin{abstract}
We describe an approximate solution to the Heisenberg operator equations of
motion for an atom in a laser field. The solution is based on a quantum
generalization of the physical picture given by the well-known Simple Man Model (SMM).
We provide  justification of the plausibility of this generalization and test its
validity by applying it for
the calculation of the coordinate and velocity autocorrelation functions which, 
to our knowledge,
have not been studied before in the context of the strong field ionization. 
Both our model and results of the {\it ab initio} numerical calculations show 
distinct types of correlations due to different types of electron's motion 
providing a useful insight into the 
strong field ionization dynamics.
\end{abstract}

\pacs{32.80.Rm 32.80.Fb 42.50.Hz}
\maketitle

\section*{Introduction}

An atom exposed to a strong laser field can be ionized. 
The foundation of the quantum theory describing this process in strong fields has been 
laid out in the seminal paper by Keldysh \cite{Keldysh64} (also known as 
the strong field approximation or the SFA theory). The Keldysh  theory introduces
the well-known classification of the ionization 
phenomena based on the value of the Keldysh parameter
$\gamma=\omega\sqrt{2|\varepsilon_0|}/E$
($\omega$, $E_0$ and $|\varepsilon_0|$ are the frequency, field strength and ionization potential
of the target system expressed in atomic units). The ionization regime corresponding to the values
$\gamma >>1 $ is known as the multi-photon regime.
The opposite limit $\gamma \lesssim 1$ is known as the tunneling regime \cite{kri}. Depending on the ionization regimes, 
the ionization process is described in drastically different ways 
\cite{Keldysh64,kri}.

The tunneling regime is particularly interesting since many interesting and important phenomena 
occurring in this regime,  such as the high harmonic generation (HHG), the attosecond pulse generation and the 
above threshold ionization (ATI), can be understood using fairly simple physical picture, the so-called 
simple man model (SMM)\cite{hhgd,Co94,kri,tipis,arbm}. In the framework of this model the electron's motion after the 
ionization event is described using classical equations of motion for an electron in the
presence of the laser field, neglecting the effect of the atomic potential. 
These equations can be easily solved leading to the testable
predictions (such as, e.g., the maximum energy of the direct electrons in the ATI process, or the cutoff photon 
energy for the HHG process) 
which often agree remarkably well with results of the {\it ab initio} quantum simulations. 

For the case of the electron's motion in the electromagnetic field in the absence of 
any other forces, not only the classical equations of motion but also
their quantum counterpart- the Heisenberg operator equations of motion for the operators of 
coordinate and momentum, can be solved fairly easily. Indeed, the solutions are practically identical.
The question arises can this fact be somehow exploited? The great utility of the SMM, on one hand,
and the fairly simple expressions for the quantum solutions to the Heisenberg 
equations of motion for the SMM physical settings suggest that we may try to somehow 
extend the SMM into the quantum domain. 

\section*{Theory}

\begin{figure}[h]
\includegraphics[width=0.9\textwidth]{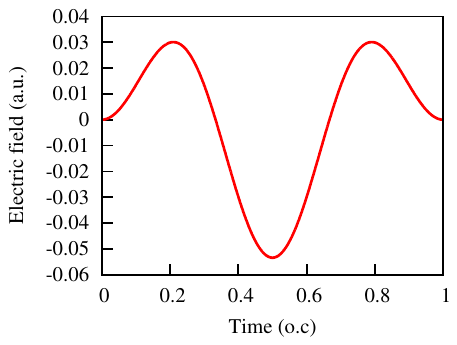}
\caption{(Color online) Electric field of the pulse \eref{ef} with $E_0=0.0534$ a.u.}
\label{fig0}
\end{figure}

We consider a hydrogen atom interacting with the laser pulse. In the Schr\"odinger picture the
system evolves according to:

\begin{equation}
i {\partial \Psi(\r,t) \over \partial t}=
\left(\hat H_0 + \hat H_{\rm int}(t)\right)
\Psi(\r,t) \ ,
\label{tdse}
\end{equation}

where $\hat H_0={\hat{\p}/2}-1/r$, and we use velocity form  
$\hat H_{\rm int}(t)= {\bm A(t)\cdot \hat{\p}} + {\bm A(t)}^2/2 $ 
for the interaction operator. Initial state of the system is $\Psi(\r,0)=\phi_0(\r)$- the ground state  of the 
hydrogen atom. 
Laser pulse is linearly polarized (along the $z-$ axis) and is defined by the vector potential:

\begin{equation}
{\bm A(t)}= -\hat {\bm z} {E_0\over \omega}\sin^2{\left\{\pi t\over T_1\right\}}\sin{\omega t} \ ,
\label{ef}
\end{equation}

with peak field strength  $E_0$, carrier frequency $\omega$, and total duration $T_1=T$, where 
$T=2\pi/\omega$ is an optical cycle (o.c.) corresponding to the frequency $\omega$.
We use the dipole approximation to describe atom-field interaction, so the expression \eref{ef} 
for the vector potential does not contain the spatial variables.
We will consider below the pulses 
with various peak field strengths $E_0$. 
We use a short pulse of one o.c total duration, so that electric field of the pulse has 
a well-defined global
maximum to facilitate study of the ionization dynamics. 
\Fref{fig0} shows the shape of the pulse \eref{ef} for 
$E_0=0.0534$ a.u. (intensity of $10^{14}$ W/cm$^2$). 

For the reader's convenience we recapitulate first some well-known facts and introduce 
few definitions which we will use below. \Eref{tdse} describes ionization process in the 
Schr\"odinger picture. In the Heisenberg picture the wave-function does not evolve in time while
the operators evolve according to:

\be
\hat Q_H(t)= \hat U(0,t) \hat Q \hat U(t,0) \ ,
\label{hp}
\ee

where $\hat Q$ stands for either coordinate or momentum operator, and 
$\hat U(t,0)$ is the time-evolution operator. In the following we will 
reserve the subscript $H$ for the operators in the Heisenberg picture, operators
without subscripts will correspond to the Schro\"odinger picture. We will use also the 
time-dependent operators $\hat Q_0(t)$ and $\hat Q_V(t)$ which we define as:

\begin{eqnarray}
\hat Q_0(t)=& \hat U_0(0,t) \hat Q \hat U_0(t,0) \nonumber \\
\hat Q_V(t)=& \hat U_V(0,t) \hat Q \hat U_V(t,0) \ ,  \nonumber \\
\label{hv}
\end{eqnarray}

where 
$\displaystyle  \hat U_0(t,0) = \exp{\left\{-i\hat H_0 t \right\}}$ is the field-free 
atomic evolution operator, and

\be
\hat U_V(t,0)= \exp{\left\{ -{i\over 2} \int\limits_{0}^t (\hat \p +\A(x))^2\ dx \right\}} 
\label{vp}
\ee

is the so-called Volkov time-evolution operator \cite{tunr2,kri} (hence the subscript 'V' in 
\Eref{hv}). This time-evolution operator describes evolution driven by the Volkov Hamiltonian 
$\hat H_V(t)=  \hat T + \hat H_{\rm int}(t)$ ($\hat T$ is the kinetic energy operator) 
of a free electron in the field of the pulse \eref{ef}.
We assume that both $\hat Q_0$ and $\hat Q_V$ can be calculated numerically 
or analytically.  Introducing
a complete set of the eigenstates $|n\rangle$ of $\hat H_0$, one can write for $\hat Q_0(t)$:

\be
\hat Q_0(t)= \isum\limits_{n,m} \langle n|\hat Q|m\rangle 
e^{i(\varepsilon_n-\varepsilon_m)t} |n\rangle\langle m| \ ,
\label{q0}
\ee

where summations run over the spectrum (including the continuous spectrum) of $\hat H_0$ with eigenenergies $\varepsilon_n$. 
As far as $\hat Q_V(t)$ is concerned, the Heisenberg operator equations of motion can
easily be solved for the case of the Volkov Hamiltonian, giving for the physically interesting
cases of the coordinate and kinetic momentum operators the simple expressions:

\begin{eqnarray}
\hat \r_V(t)&=& \r + {\hat \p} t + \int\limits_0^t {\bm A}(\tau)\ d\tau  \nonumber \\  
\hat \ve_V(t)&=& {\hat \p} + {\bm A}(t) \ , \nonumber \\ 
\label{hv1}
\end{eqnarray}

where $\r$ and $\hat{\p}$ are the Schr\"odinger coordinate and momentum operators. In the atomic 
units system we use the kinetic momentum coincides with the velocity. We employed, therefore,
the notation ${\hat \ve}_V(t)$ in \eref{hv1}. We would like to remind also that we 
use the velocity gauge to describe atom-field interaction, hence presence of 
the vector potential in the second equation \eref{hv1}. 

Equations \eref{hv1} look exactly like their
classical counterparts used in the simple man model. This fact motivated us to try to incorporate
them into a quantum version of the SMM, relying on the Heisenberg picture. The exact Heisenberg
equations of motion for operators $\hat{\r}_H(t)$ and $\hat{\ve}_H(t)$
following from the definitions \eref{hp}:

\begin{eqnarray}
{d\over dt} {\hat{\r}}_H(t)&=& \hat{\ve}_H(t)  \nonumber \\
{d\over dt} {\hat{\ve}}_H(t)&=& -i\left[\hat{\p}_H(t),{\hat V}({\hat{\r}}_H(t))\right] +
{d\over dt} {\bm A}(t) \ , \nonumber \\
\label{he}
\end{eqnarray}

(here ${\hat V}({\hat{\r}}_H(t))= -1/|{\hat{\r}}_H(t)|$ for hydrogen atom, 
$[\hat A,\hat B]=\hat A\hat B-\hat B\hat A$ denotes
the commutator of operators $\hat A$ and $\hat B$, and  
$\hat \ve_H(t)= {\hat \p}_H(t) + {\bm A}(t) $  are, of course, too complicated to be solved
exactly (or even numerically). We can try, however, to find an approximate solution using 
the time-dependent coordinate and momentum operators we introduced above 
in \Eref{q0} and \Eref{hv1} and the physical insight we get from the classical SMM.

The predictive power of the SMM shows that the simple expressions for the 
classical electron's coordinate and velocity in the laser field which neglect 
all atomic interactions, 
can be used with success to explain many an ionization phenomena. It is natural, therefore, to
use the quantum counterpart of these classical expressions given by the \Eref{hv1} in trying to 
construct an approximate solution to the Heisenberg equations of motion \eref{he}. In the limit of the
vanishing field the solution should, of course, reduce to the field-free operators
${\hat{\r}}_0(t)$ and ${\hat{\p}}_0(t)$ obtained by substituting the Schr\"odinger coordinate and momentum 
operators, respectively, for the operator $\hat Q$ in \Eref{q0}. This reasoning suggests
the following tentative expression for the approximate solution to the Heisenberg equations of
motion for the coordinate operator:

\be
{\hat{\r}}_H(t)= {\hat{\r}}_0(t) + \alpha(t) \left(\r + {\hat \p}t\right) + {\bar\r}(t) \ .
\label{tr}
\ee

On the right-hand side of this equation ${\hat{\r}}_0(t)$ is the field-free 
Heisenberg coordinate operator obtained using \Eref{q0}, 
$\r$ and ${\hat \p}$ are time-independent Schr\"odinger operators, and we absorbed the $c$-number term
in the first of the equations \Eref{hv1} into the $c$-number term ${\bar\r}(t)$ in 
\Eref{tr}. That this term coincides with the expectation value of the coordinate can be easily seen from the
\Eref{tr} by noting that in the Heisenberg picture this expectation value is just the expectation value of 
the operator ${\hat{\r}}_H(t)$ obtained using the wave-function of the ground state of atomic hydrogen.
The expectation values 
of ${\hat{\r}}_0(t)$, $\r$ and ${\hat \p}$ in this state are all zero. The real function $\alpha(t)$ 
(it must be real to preserve hermicity of the coordinate operator) 
in \Eref{tr} is yet unknown. Basing on the general structure of the \Eref{tr} and the simple physical
picture of ionization provided by the SMM, we may say that the first term on the r.h.s. of
the \Eref{tr} describes atomic and the second term describes ionized electrons. 
The function $\alpha(t)$ then must be related to the ionization probability. This assumption 
can be justified by the following reasoning. 

Using the definition for the Heisenberg coordinate operator and the Dyson equation for the 
evolution operator $\hat U(t,0)$,
corresponding to the 
partition $\hat H= \hat H_0+ \hat H_{\rm int}(t)$ of the total Hamiltonian:

\be
\hat U(t,0)= \hat U_0(t,0) -i\int\limits_0^t \hat U(t,\tau) \hat H_{\rm int}(\tau)\hat U_0(\tau,0)\ d\tau \ ,
\label{dy}
\ee

one obtains the following approximate equation for the Heisenberg coordinate operator: 

\begin{eqnarray}
{\hat{\r}}_H(t) &\approx &  \hat{\r}_0(t) \nonumber \\  
&+& \left [ -i \hat U_0(0,t)\r\int\limits_0^t \hat U(t,\tau) \hat H_{\rm int}(\tau)\hat U_0(\tau,0)\ d\tau + 
h.c\right] \ , \nonumber \\
\label{d1}
\end{eqnarray}

where h.c. here and below stands for the Hermitian conjugate.
In deriving \Eref{d1} we neglected the term quadratic in $\hat H_{\rm int}(\tau)$ originating when
the r.h.s of \Eref{dy} is substituted into the equation defining time-dependent operators in the 
Heisenberg picture. Since the second term on the r.h.s. of the Dyson equation \eref{dy} is related to the
ionization amplitude and is small for the not too strong fields which we consider below,
the neglect of the terms quadratic in $\hat H_{\rm int}(\tau)$
is legitimate. From the Dyson equation written in a different form (which corresponds to the 
partition $\hat H= \hat H_V+ \hat V_{\rm atom}$ of the total Hamiltonian):

\be
\hat U(t,0)= \hat U_V(t,0) -i\int\limits_0^t \hat U(t,\tau) \hat V_{\rm atom} \hat U_0(\tau,0)\ d\tau \ ,
\label{dyppt}
\ee

where $V_{\rm atom}$ is the  
atomic potential, one can see that 
replacing the operator $\hat U_0(0,t)$ on the r.h.s of the 
\Eref{d1} with $\hat U_V(0,t)$ would result in additional terms 
bilinear in $\hat H_{\rm int}(\tau)$ and $\hat V_{\rm atom}$.
Assuming again that these terms can be neglected comparing to the leading term, we can write:

\begin{eqnarray}
{\hat{\r}}_H(t) &\approx &  \hat{\r}_0(t) \nonumber \\  
&+& \left [ -i \hat U_V(0,t)\r\int\limits_0^t \hat U(t,\tau) \hat H_{\rm int}(\tau)\hat U_0(\tau,0)\ d\tau + 
h.c\right] \ . \nonumber \\
\label{d2}
\end{eqnarray}

Consider the combination of the operators $\displaystyle \hat U_V(0,t)\r \hat U(t,\tau)$ 
on the r.h.s of the \Eref{d2}. We can write for this combination:

\begin{eqnarray}
\hat U_V(0,t)\r\hat U(t,\tau)&=& \hat U_V(0,t)\r\hat U_V(t,0)\hat U_V(0,t)\hat U(t,\tau) \nonumber \\
&=& \hat{\r}_V(t) \hat U_V(0,t)\hat U(t,\tau) \nonumber \\
&\approx& \hat{\r}_V(t) \hat U(0,t)\hat U(t,\tau) \nonumber \\
&= & \hat{\r}_V(t) \hat U(t,\tau) \ ,  \nonumber \\
\label{d3}
\end{eqnarray}

where we used multiplication properties 
(such as $\displaystyle  \hat U(t_1,t_2)\hat U(t_2,t_3)=\hat U(t_1,t_3)$) for the evolution operators 
$\hat U(t_1,t_2)$ and $\hat U_V(t_1,t_2)$ and used the identity
$\hat{\r}_V(t)=\hat U_V(0,t)\r\hat U_V(t,0)$, where $\hat{\r}_V(t)$ is the Volkov coordinate
operator introduced by \Eref{hv1}.
Furthermore, in deriving \Eref{d3} we replaced in the third line 
the Volkov propagator $\hat U_V(0,t)$ with the total propagator $\hat U(0,t)$. This is the
replacement analogous to the one usually done in the derivation of the Keldysh expression for the ionization 
amplitude in the framework of the SFA (by replacing $\hat U(0,t)$ on the r.h.s of the \Eref{dy} with Volkov 
propagator $\hat U_V(0,t)$ one can obtain the well-known expression for the Keldysh ionization amplitude).
The justification of this
replacement is again the assumption, which we have been using systematically, 
that the terms of the higher order 
(quadratic in $\hat H_{\rm int}(\tau)$ in this instance) 
can be neglected comparing to the leading term. From \Eref{d2} and \Eref{d3} we obtain:
 
\begin{eqnarray}
{\hat{\r}}_H(t) &\approx &  \hat{\r}_0(t) \nonumber \\  
&+& \left [ -i \hat{\r}_V(t)\int\limits_0^t \hat U(0,\tau) \hat H_{\rm int}(\tau)\hat U_0(\tau,0)\ d\tau + 
h.c\right] \ . \nonumber \\
\label{d4}
\end{eqnarray}

Introducing the complete set $|n\rangle$ of the eigenstates of the 
field-free Hamiltonian, 
the operator expression under the integral sign in \Eref{d4} can be written as:

\begin{eqnarray}
\hat U(0,\tau) \hat H_{\rm int}(\tau)\hat U_0(\tau,0) = \nonumber \\ 
\isum_{n,m} \langle n|\hat U(0,\tau) \hat H_{\rm int}(\tau)|\hat U_0(\tau,0)|m\rangle\ |n\rangle\langle m| 
= \nonumber \\
\isum_{n,m} \langle \Psi_n(\tau)|\hat H_{\rm int}(\tau)|\phi_m(\tau)\rangle \ |n\rangle\langle m|= \nonumber \\
-i{\partial \over \partial\tau} \isum_{n,m} \langle \Psi_n(\tau)|\phi_m(\tau)\rangle \ |n\rangle\langle m|\ ,\nonumber \\
\label{d5}
\end{eqnarray}

where $\Psi_n(\tau)=\hat U(\tau,0)\phi_n$, $\phi_m(\tau)=\hat U_0(\tau,0)\phi_m=e^{-i\varepsilon_m\tau}\phi_m$,
$\phi_m$ and $\varepsilon_m$ are, respectively, the eigenstates and eigenenergies of the field-free atomic
Hamiltonian. The overlaps $T_{nm}=\langle \Psi_n(\tau)|\phi_m(\tau)\rangle$ form the matrix of transition
amplitudes between various field-free states of the atomic Hamiltonian. Substituting \Eref{d5} into
\Eref{d4} we obtain:

\be
{\hat{\r}}_H(t) \approx  \hat{\r}_0(t) + \left [ \hat{\r}_V(t) \hat T + h.c\right] \ ,
\label{d6}
\ee

where the transition operator $\hat T$ is given by:

\be
\hat T= -\isum_{n,m} 
\left( \langle \Psi_n(t)|\phi_m(t)\rangle - \delta^n_m \right) |n\rangle\langle m|
\label{d7}
\ee

\Eref{d6} is still pretty complicated. To advance further we have to make an assumption   
about the operator $\hat T$ in \Eref{d7}. This operator is clearly related to the 
transitions and ionization probabilities, becoming a zero operator in the absence of the
electric field. Having in mind that we are developing a 
model based on the simple physical picture we will replace this operator with 
a $c-$number function which, taking into account the properties of the operator 
$\hat T$, can be considered proportional to the ionization probability. The Heisenberg
operator ${\hat{\r}}_H(t)$ obtained in this way is then a sum of two terms, the 
field-free atomic time-dependent operator $\hat{\r}_0(t) $ and the time-dependent Volkov
coordinate operator $\hat{\r}_V(t)$ entering the expression with the weight proportional to the 
ionization probability. Substituting the expression \eref{hv1} for $\hat{\r}_V(t)$, 
we obtain:

\be
{\hat{\r}}_H(t) \approx  \hat{\r}_0(t) + 
\alpha(t)\left(\r + {\hat \p} t + \int\limits_0^t {\bm A}(\tau)\ d\tau \right) \ ,
\label{d8}
\ee

where $\hat{\r}_0(t) $ is the field-free atomic time-dependent coordinate operator
which can be obtained using \Eref{q0}, $\r$ and $\p$ are usual coordinate and
momentum operators in the Schro\"odinger representation, ${\bm A}$- the vector potential, and
$\alpha(t)$ is a function which we assume to be proportional to the ionization probability.
Taking into account that expectation value of the 
coordinate operator in the Heisenberg picture is just the matrix element 
$\displaystyle \langle \phi_0|{\hat{\r}}_H(t)|\rangle \phi_0$ (where $\phi_0$ is the initial
state of the system which is the ground state of hydrogen in our case), and 
that expectation values of both  $\hat{\r}_0(t) $, $\r$ and ${\hat \p}$ operators vanish in
the ground state of hydrogen, we may absorb the $c$-number term in \Eref{d8} into the 
expectation value, which gives us the \Eref{tr} for the coordinate operator in the 
Heisenberg representation.

The expression for the the velocity operator 
$\hat{\ve}_H(t)$ follows from \Eref{tr} by the time differentiation as indicated in the first of
the \Eref{he}.  The canonical momentum can then be found as $\hat \p_H(t)=\hat{\ve}_H(t) - {\bm A}(t)$. 
We note that while this procedure gives us the Hermitian  Heisenberg coordinate
and momentum operators, it does not preserve the correct commutation relations 
$[\hat p^{(n)}_H(t),\hat r^{(m)}_H(t)]=-i\delta^m_n$,\,
This is a consequence of the fact that the 
transformation of the Schro\"odinger pair of operators $\r$, $\hat \p$ to the  Heisenberg pair 
$\hat \r_H(t)$, $\hat \p_H(t)$ described by the \Eref{tr} is not unitary. The lack of unitarity,
however, should not be considered as a serious drawback. Many widely used approaches to the description 
of the ionization in strong fields, e.g., the SFA approach have a similar problem. 
The SFA is based on the Schro\"odinger picture and the non-unitarity of this method manifests 
itself as a non-unitary evolution of the 
state vector \cite{sfam}. Consequently, the total sum of all the probabilities is not conserved 
and may not sum up to unity
in the SFA. The great success and utility of the SFA show, however, this of unitarity
of the method does not constitute a major impediment. 
We will study below some consequences and testable predictions we may derive from \Eref{tr}.

\section*{Results}

An advantage which the Heisenberg picture offers, is the natural 
way in which the many-time correlation functions, containing detailed 
information about evolution of the system, can be introduced. We will be
interested below in the two-time autocorrelation functions, defined
as follows:

\begin{eqnarray}
C_{QQ}(t_2,t_1)&=& \langle \phi_0|\left(\hat Q(t_2)-\bar Q(t_2)\right) 
\left(\hat Q(t_1)-\bar Q(t_1)\right)|\phi_0\rangle \nonumber \\
&=& \langle \phi_0|\hat Q(t_2)\hat Q(t_1)|\phi_0\rangle - 
\bar Q(t_2)\bar Q(t_1) \ , \nonumber \\
\label{cor}
\end{eqnarray} 

where $\phi_0$ is the ground state of the hydrogen atom, $\hat Q(t)$
is an operator in the Heisenberg picture and $\bar Q(t)$ is the 
expectation value of  $\hat Q(t)$. Since we consider only  
the ground state of hydrogen as the initial state in the present work, we will omit below $\phi_0$ in the formulas, implicitly understanding
that $\langle \hat A \rangle=\langle\phi_0|\hat A |\phi_0\rangle $ for
any operator $\hat A$. 
We will consider below as two examples
the autocorrelation functions  $C_{zz}(t_2,t_1)$ and
$C_{v_z,v_z}(t_2,t_1)$ for the components of the coordinate and velocity 
operators in the direction of the laser field. 

On one hand, we can compute the autocorrelation functions
in an {\it ab initio} way  
without any approximations using the well-tested procedure \cite{cuspm} 
we use to solve the time-dependent Schr\"odinger equation (TDSE).
Considering $C_{zz}(t_2,t_1)$ as an example, using definition 
\eref{hp},  and employing the well-known properties of the
time-evolution operators, we may write:

\be
\langle \hat z_H(t_2)\hat z_H(t_1)\rangle =  
\langle\hat U(t_2,0)\phi_0|z\hat U(t_2,t_1)z|\hat U(t_1,0)\phi_0\rangle.
\label{tdcor}
\ee

\begin{figure}[h]
\includegraphics[width=0.9\textwidth]{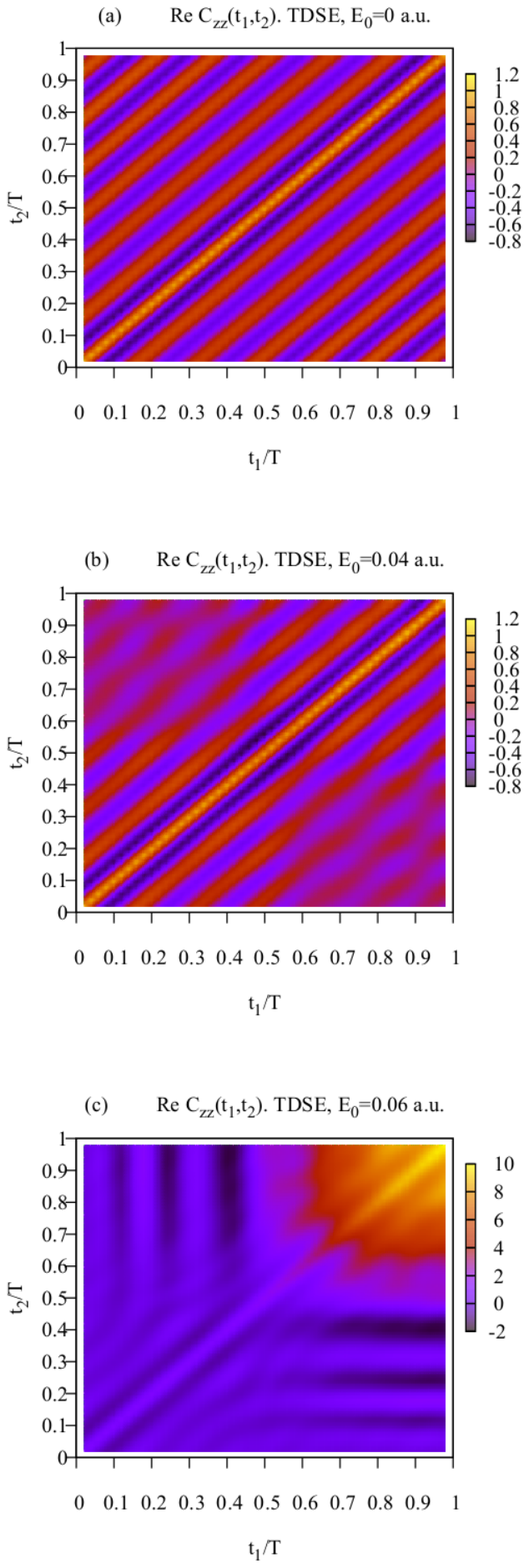}
\caption{(Color online) Real part of the coordinate correlation function $C_{zz}(t_1,t_2)$ 
obtained  from  the TDSE calculation using \Eref{tdcor}. {\bf a} Peak field strength $E_0=0$ a.u.
{\bf b} $E_0=0.04$ a.u. {\bf c} $E_0=0.06$ a.u.}
\label{tdsef}
\end{figure}

\begin{figure}[h]
\includegraphics[width=0.9\textwidth]{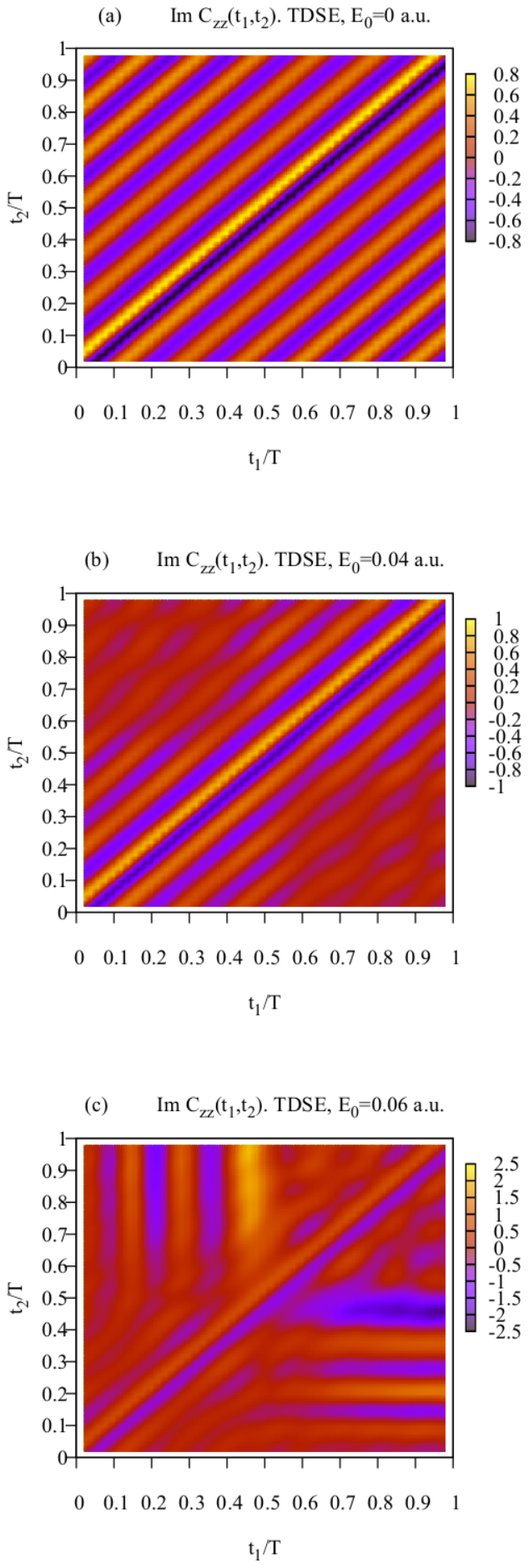}
\caption{(Color online) Imaginary part of the 
coordinate correlation function $C_{zz}(t_1,t_2)$ 
obtained  from  the TDSE calculation using \Eref{tdcor}.
{\bf a} $E_0=0$ a.u.
{\bf b} $E_0=0.04$ a.u. {\bf c} $E_0=0.06$ a.u.}
\label{tdsefi}
\end{figure}

Calculation of this expression requires, thus, first propagating the 
TDSE starting with $\Psi(0)=\phi_0$- the 
ground state of hydrogen, on the interval
$(0,t_1)$, thus obtaining the state vector  $\Psi(t_1)$ at $t=t_1$.
Acting with the (Schr\"odinger) operator $z$ on this vector we obtain the
wave-function $\Psi_1(t_1)=z\Psi(t_1)$. $\Psi_1(t_1)$ is further
propagated on the interval 
$(t_1,t_2)$ yielding the wave-function $\Psi_1(t_2)$, from which we
obtain $\Psi_2(t_2)=z\Psi_1(t_2)$. Finally, 
the autocorrelation function can be found by projecting  $\Psi_2(t_2)$
on the state vector $\Psi(t_2)$, obtained by solving the TDSE with the 
initial condition $\Psi(0)=\phi_0$ on the interval
$(0,t_2)$. These calculations were performed using
the numerical procedure \cite{cuspm} for the solution of the 
TDSE for hydrogen atom in presence of the  
laser field given by \Eref{ef}. Results of the {\it ab initio} 
TDSE calculations of the real and imaginary parts of $C_{zz}(t_1,t_2)$ for different field strengths 
are shown in \Fref{tdsef}, \Fref{tdsefi}.

The autocorrelation functions 
$C_{zz}(t_1,t_2)$  shown in the \Fref{tdsef}, \Fref{tdsefi}  
show two distinctly different types of correlations. 
The correlation patterns for small fields- the diagonal stripes in the \Fref{tdsef}
\Fref{tdsefi} are reminiscent of those 
we would obtain for  the field-free Heisenberg operators. The origin
of the stripes is clear from the expression \Eref{q0}, from
which we can obtain:

\be
C_{zz}(t_1,t_2)= \isum\limits_{m} |\langle 0|\hat Q|m\rangle|^2 
e^{i(\varepsilon_0-\varepsilon_m)(t_1-t_2)} \ ,
\label{q01}
\ee
  	
where $|0\rangle$ is the hydrogen ground state. We can also explain the
correlations pattern in the \Fref{tdsef}, \Fref{tdsefi} qualitatively using a simple classical picture of
the atomic periodic motion.  Electron's positions at the moments
$t$ and $t+kT_a$, where $T_a$ is the period of the electron's orbital motion and $k$ is an integer, 
are clearly correlated, analogously, the electron's positions 
at the moments $t$ and $t+kT_a/2$ are anticorrelated, hence the pattern
of the maxima and minima of the autocorrelation function running diagonally.
Strictly speaking, this simple classical explanation is 
applicable to the hydrogen atom with some reservations, since the different terms in the 
\Eref{q0} are not equally spaced in energy. Consequently the frequencies in \Eref{q01} are not 
integer multiples of some base frequency, as the simple classical picture we alluded to above 
implies. Had we considered instead of the hydrogen atom the  
harmonic oscillator with frequency $\Omega$ and equally spaced energy levels, \Eref{q0} would result in 
the well-known relation for the coordinate operator in the Heisenberg picture 
$\hat z_H(t)= z\cos{\Omega t}+ {\hat p}_z/\Omega\sin{\Omega t}$ \cite{mess}. For such
$\hat z_H(t)$ we would have obtained
a pattern of equally spaced maxima and minima of the autocorrelation
function running diagonally, for which the classical explanation would be perfectly adequate. 
Nevertheless, we shall use this line of arguments based on the picture of the 
electron's periodic motion for
the purposes of the qualitative analysis even for hydrogen, having 
in mind that for this system this picture may have some limitations.      
In classical terms, thus,  the pattern of correlations in the field-free case
just reflects the periodic orbital motion of the electron. 
For non-zero external fields this field-free pattern of correlations
becomes first perturbed (for $E_0=0.04$ a.u.) and then is almost 
completely superseded by a different pattern for $E_0=0.06$ a.u., which exhibits 
horizontal and vertical stripes rather than the diagonal ones, and
which shows a good deal of correlated motion for both $t_1$, $t_2$ near the
end of the pulse. We shall see below that this new pattern of correlations induced by the field can
be explained in considerable detail by our model based on the 
\Eref{tr}.  
     		
Using 
\Eref{tr} and \Eref{cor} we obtain for the  
autocorrelation function  $C_{zz}(t_2,t_1)$:

\begin{eqnarray}
C_{zz}(t_2,t_1) & = &  C^0_{zz}(t_2,t_1) \nonumber \\
&+ & \alpha(t_1)\langle{\hat{z}}_0(t_2)\left(z + {\hat p_z}t_1\right)\rangle + 
\alpha(t_2)\langle{\hat{z}}_0(t_1)\left(z + {\hat p_z}t_2\right)\rangle \nonumber \\ 
&+ & \alpha(t_1)\alpha(t_2)\langle \left(z + {\hat p_z}t_1\right)
\left(z + {\hat p_z}t_2\right)\rangle \ , \nonumber \\ 
\label{mm}
\end{eqnarray}

where $C^0_{zz}(t_2,t_1)$ is the field-free autocorrelation function, 
$\hat z_0(t)$ is the $z-$ component of the 
the field-free Heisenberg coordinate operator obtained using \Eref{q0}, 
$z$ and ${p_z}$ are the usual time-independent Schr\"odinger operators
of the $z-$ components of coordinate and momentum. 
Calculations of the expectation values appearing in the 
second line of \Eref{mm} can be done as follows. 
Considering the product $\hat z_0(t_2)\hat p_z$ as an example, we
may write using the definition of $\hat z_0(t)$: 
$\displaystyle 
\langle \hat z_0(t_2)\hat p_z\rangle = 
e^{i\varepsilon t_2}\langle z\phi_0|\hat U_0(t_2,0)|\hat p_z\phi_0\rangle$,
where $\hat U_0(t_2,0)$ is the field-free atomic evolution operator. Calculation
of this matrix element requires thus field-free propagation of the 
initial state $\hat p_z\phi_0$ on the interval $(0,t_2)$, which can be 
done without difficulties using the numerical procedure we use to solve the
TDSE. 
Calculations of the 
expectation values in the third line of \Eref{mm} does not pose
any difficulties.  
There is one more ingredient which we have to provide to use  \Eref{mm}, 
it is the function $\alpha(t)$. As we have noted above, 
the reasoning based on the physical picture of the ionization process suggests that 
$\alpha(t)$ should be related to the ionization probability $P(t)$. We will
use, therefore, the expression $\alpha(t)=cP(t)$ for the function
$\alpha(t)$ in \Eref{tr} and \Eref{mm}, where $P(t)$ is the ionization
probability, and $c$ is a constant factor. We compute $P(t)$ as
$\displaystyle P(t)=\int\limits_0^t W_{YI}(\tau)\ d\tau$, where
$W_{YI}(\tau)$ is the well-known Yudin-Ivanov instantaneous
ionization rate (YI IIR) \cite{yi}. Our model, thus, has one
free parameter $c$ which we can vary to achieve better agreement 
between the  {\it ab initio} $C^0_{zz}(t_2,t_1)$, and the 
$C^0_{zz}(t_2,t_1)$ we obtain from \Eref{mm}.

\begin{figure}[h]
\includegraphics[width=0.9\textwidth]{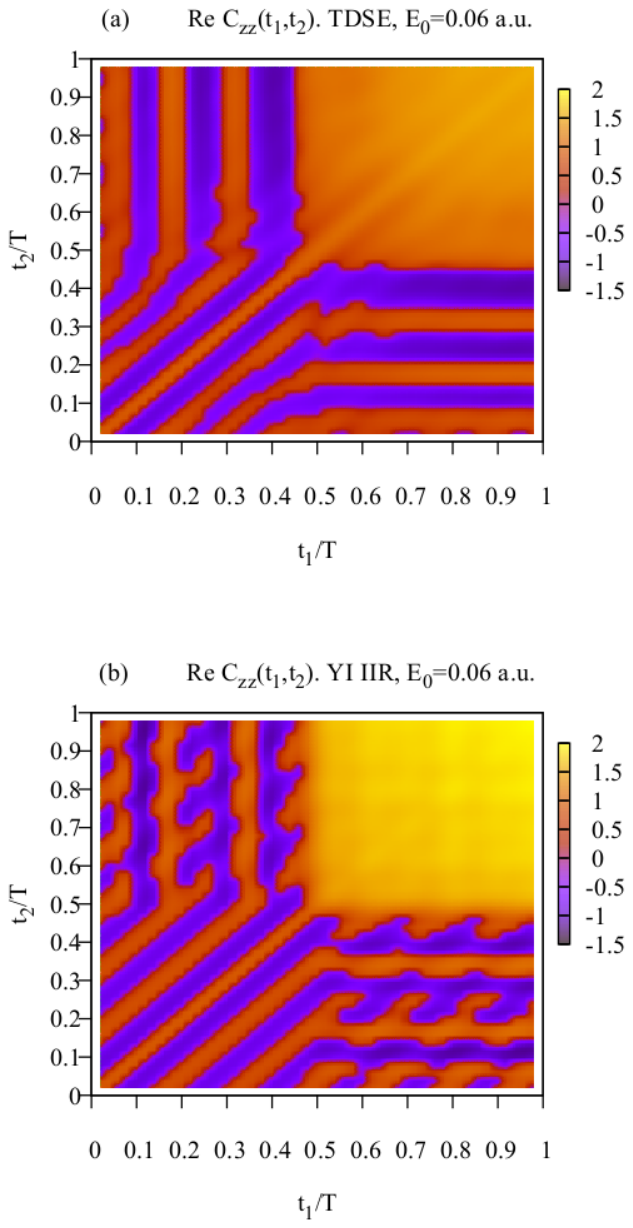}
\caption{(Color online) Real part of the 
coordinate correlation function \eref{cor}.  {\bf a} Results of the TDSE calculation
(\Eref{tdcor}). {\bf b} Results obtained using \Eref{mm} with $\alpha(t)$ computed as 
ionization probability using Yudin-Ivanov ionization rate (YU IIR). For better visibility and to reveal 
more detail we plot the quantity $(\rm{Re}(C_{zz}(t_1,t_2))^{1/5}$. 
}
\label{yiz}
\end{figure}

\begin{figure}[h]
\includegraphics[width=0.9\textwidth]{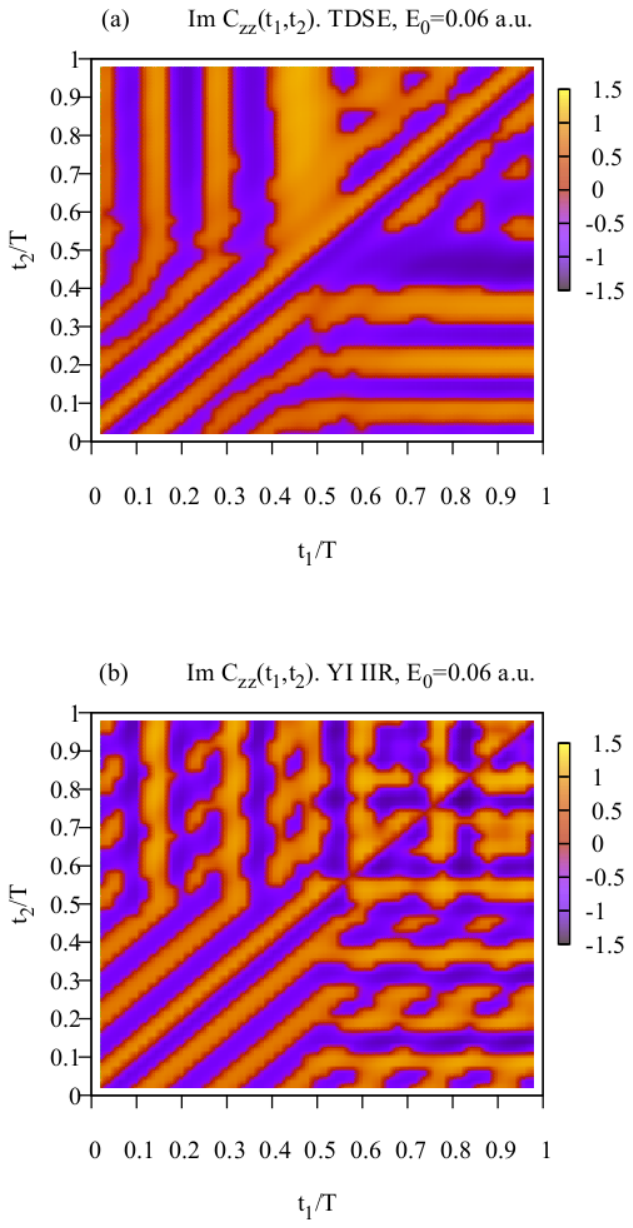}
\caption{(Color online) Imaginary part of the 
coordinate correlation function \eref{cor}. {\bf a} Results of the TDSE calculation
(\Eref{tdcor}). {\bf b} Results obtained using \Eref{mm} with $\alpha(t)$ computed as 
ionization probability using Yudin-Ivanov ionization rate (YU IIR). 
The quantity $(\rm{Im}(C_{zz}(t_1,t_2))^{1/5}$ is plotted.
}
\label{yizi}
\end{figure}

A comparison of the results we obtain in this way, using our model
and the {\sl ab initio} TDSE results, is shown in \Fref{yiz}, \Fref{yizi} . 
To reveal more detail we use a nonlinear scale in the Figures, presenting
fractional power of $\rm{Re}(C_{zz}(t_1,t_2))$ and
 $\rm{Im}(C_{zz}(t_1,t_2))$. We can see from \Fref{yiz}, \Fref{yizi} that
both TDSE and our model exhibit a new type of correlations- the
vertical and horizontal stripes. In the classical picture to which we alluded
above,  which associated the diagonal stripes in the pattern of correlations
dominant for low fields with the correlations due to the periodic motion, this new type of correlations can be accounted for as follows. Correlations
between coordinates of the electron at the moments of time $t_1$ and 
$t_1+kT_a$ persist till the moment of time when either $t_1$ or $t_1+kT_a$ gets equal to 
the time of ionization. After the occurrence of the ionization event,
different types of correlations are introduced.
In the \Eref{mm} these types of correlations
are described by the second and third  lines of the equation.
The second line describes the 
correlations between the electrons's coordinate along the field-free 
atomic trajectory and the electron's coordinate along the ionized trajectory. These correlations
are responsible for the horizontal and vertical stripes in the
correlations pattern in \Fref{yiz}, \Fref{yizi}. Employing  the simplified classical picture 
we  used above, we might say
that this type of correlations arises because the $z$-coordinate of the 
ionized electron's wave-packet is correlated  with the electron's $z-$coordinate either at the moment of the 
electron's birth $t=t_1$, or the electron's coordinates at the moments of time $t=t_1-kT_a$ with positive
integer $k$, when electron is located at approximately the same spatial point in the course of its
orbital motion. Similarly, the minima in the correlations pattern  appear because
the $z$-coordinate of the 
ionized electron's wave-packet is anticorrelated  with the electron's $z-$coordinate 
at the moments of time $t=t_1-kT_a/2$ with positive
integer $k$, when electron is located at approximately opposite spatial point of its orbit. 
The area of highly correlated motion seen for the real part of the correlation function 
in \Fref{yiz} in the region $t_1/T>1/2$, 
$t_2/T>1/2$ is due to the term in the third line of the \Eref{mm}, which describes essentially the 
propagation of correlations for the free electron motion, 
which grows with time as $t_1t_2$ for large times.
We see that all these features are present in both the {\it ab initio} TDSE and our model calculation
based on the \Eref{mm}. Our model calculation reproduces these features quite accurately qualitatively, and
even, as the \Fref{yiz} shows,  quantitatively in the case of the real part  of the autocorrelation function. Agreement between the imaginary parts $C_{zz}(t_1,t_2)$ given by the {\it ab initio} TDSE and the model calculation based on \Eref{mm} is less spectacular but, we believe, can still be considered pretty good given that
we use only one adjustable parameter in  
our model calculations. 
That our model reproduces the real part of the 
correlation function better is, perhaps,  not surprising, taking into account  
that in applying our \Eref{mm} we 
employed only one real positive function $\alpha(t)$ proportional to the ionization probability,
discarding thereby all information about phase. Going back to the real part of the autocorrelation function, 
we would like to draw attention to the fact that the transition between the two types of
correlations which we discussed above: the field-free correlations described by the term in the first
line of the \Eref{mm}, and the 
correlations between the electrons's coordinate along the field-free atomic trajectory and the 
electron's coordinate along the ionized trajectory described by the terms in the second
line of the \Eref{mm}, is quite distinct. Given that correlation function is, in principle, an
experimentally measurable quantity \cite{cor_measure}, this offers an intriguing possibility of providing 
an experimental access to study the somewhat elusive notions such as the moment of the electron's 
birth into the continuum and
the tunneling time, which have received considerable attention in the literature lately
\cite{tor,inst1,Teeny2016,iir,landsman_bohm,nonzerot1,nonzerot2,nonzerot3,Nie,Nie1}.

\begin{figure}[h]
\includegraphics[width=0.9\textwidth]{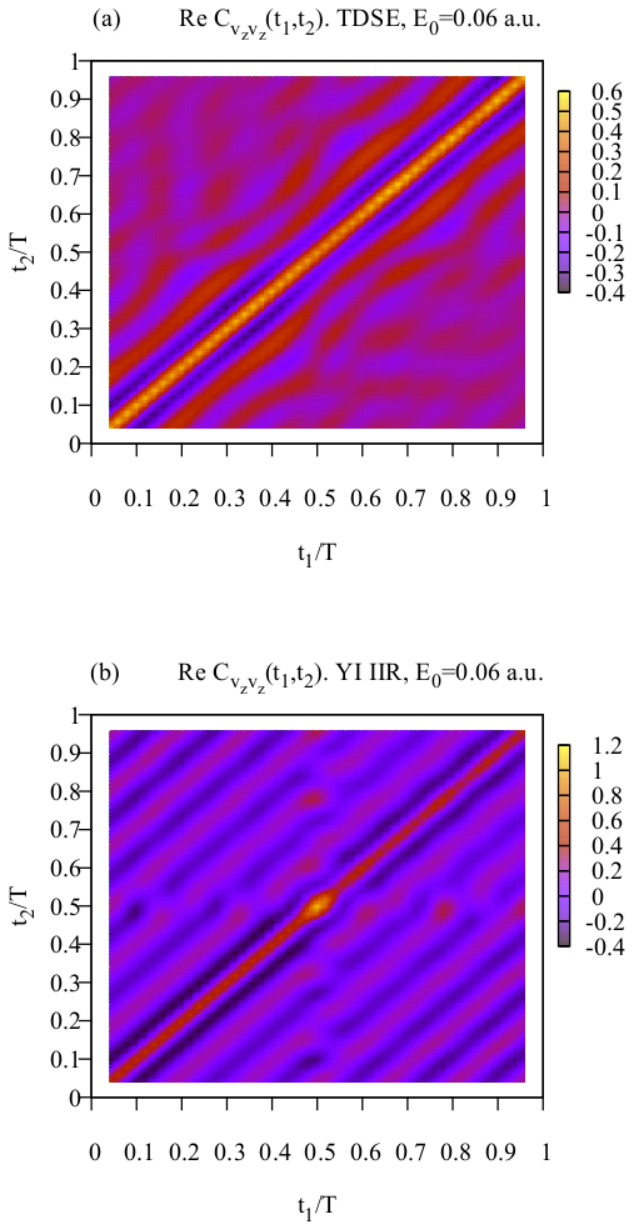}
\caption{(Color online) Real part of the 
velocity correlation function obtained using \Eref{vv}.
{\bf a} TDSE calculation
(\Eref{tdcor}). {\bf b} Results given by the \Eref{mm} with $\alpha(t)$ computed as 
ionization probability using Yudin-Ivanov ionization rate (YU IIR). 
}
\label{vvf}
\end{figure}

\begin{figure}[h]
\includegraphics[width=0.9\textwidth]{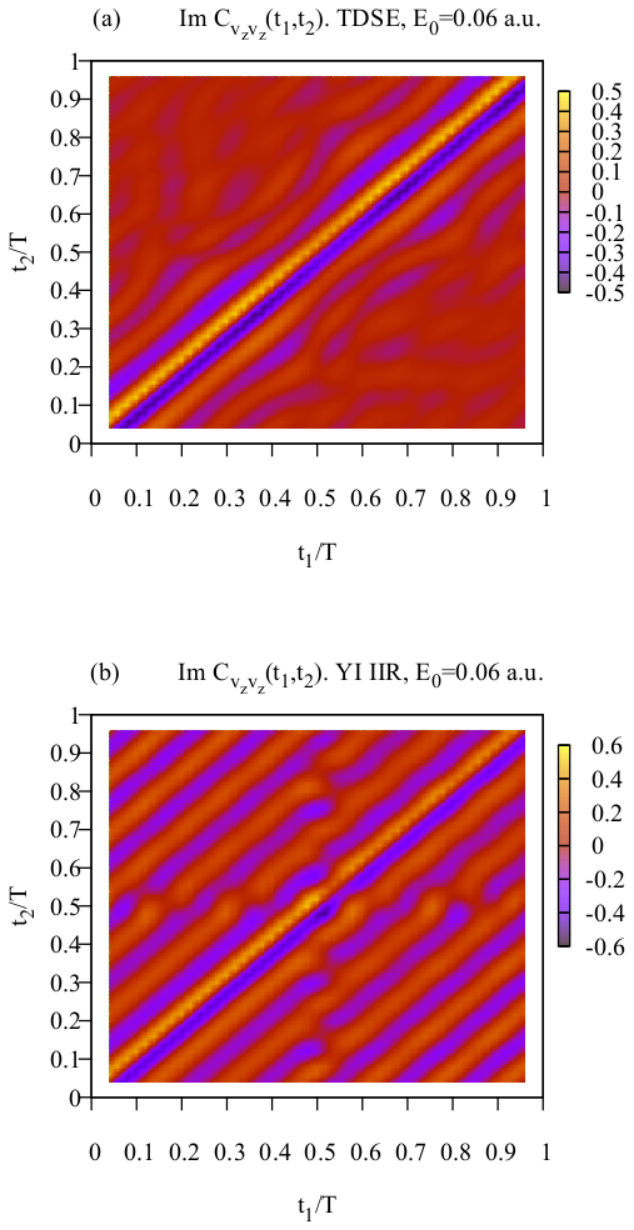}
\caption{(Color online) Imaginary part of the 
velocity correlation function obtained using \Eref{vv}.
{\bf a} Results of the TDSE calculation
(\Eref{tdcor}). {\bf b} Results given by the \Eref{mm} with $\alpha(t)$ computed as 
ionization probability using Yudin-Ivanov ionization rate (YU IIR). 
}
\label{vvfi}
\end{figure}

Having in our disposal the coordinate autocorrelation function $C_{zz}(t_1,t_2)$, we
can find other correlation functions. An example is shown in \Fref{vvf}, \Fref{vvfi} 
where we present
the velocity autocorrelation function $C_{v_zv_z}(t_1,t_2)$ obtained from $C_{zz}(t_1,t_2)$
as:

\be
C_{v_zv_z}(t_1,t_2)= {\partial\over \partial t_1}{\partial\over \partial t_2} C_{zz}(t_1,t_2) \ .
\label{vv}
\ee

The velocity correlation pattern predicted by our model based on the \Eref{vv}, \Eref{mm} agree
qualitatively well with the {\it ab initio} TDSE results.

To summarize, we presented a simple tentative solution of the Heisenberg operator equations of
motion for an atom exposed to electromagnetic field. We might call this solution a simple man model in
the Heisenberg picture since our main equation \Eref{tr} looks very much like its classical counterpart
used in the SMM.  We tested the veracity of this tentative expression by applying it to
the calculation of the coordinate and velocity autocorrelation functions.  
As the {\it ab initio}
TDSE calculations show, these functions are an interesting object of study and, to our knowledge,
have not been studied before in the context of the strong field ionization. In particular, their 
exhibiting the distinct types of correlations due to different types of electron's motion
may offer a useful insight into the physics of strong field ionization. 
We saw that the autocorrelation
functions computed by fairly simple means using  the main equation \Eref{tr} 
agree well with the results of the TDSE calculations, thereby confirming the utility of
our model.

%\begin{figure}[h]
%\begin{tabular}{ll}
%\resizebox{80mm}{!}{\epsffile{vzvz_yi_f=0.06.eps}}  &
%\resizebox{80mm}{!}{\epsffile{vzvz_td_f=0.06.eps}} \\
%\resizebox{80mm}{!}{\epsffile{vzvz_yi_f=0.08.eps}}  &
%\resizebox{80mm}{!}{\epsffile{vzvz_td_f=0.08.eps}} \\
%\resizebox{80mm}{!}{\epsffile{vzvz_yi_f=0.1.eps}}  &
%\resizebox{80mm}{!}{\epsffile{vzvz_td_f=0.1.eps}} \\
%\end{tabular}
%\caption{(Color online) Velocity autocorrelation function obtained from the TDSE and YI calculations.}
%\end{figure}

\section*{Acknowledgements}
 
This work was supported by IBS (Institute for Basic Science) under IBS-R012-D1.

\bibliography{paper}

\end{document}